\documentclass[aps,preprintnumbers,nofootinbib,superscriptaddress,11pt]{revtex4}
\usepackage[hypertex]{hyperref}
\usepackage[dvipdfmx]{graphicx} 
\usepackage{amsmath,amssymb,amsfonts,bm}

\def\a{\alpha} \def\b{\beta}   \def\d{\delta} \def\D{\Delta} \def\e{\epsilon}  \def\h{\eta} \def\th{\theta}   \def\l{\lambda} \def\L{\Lambda} \def\m{\mu} \def\n{\nu}        \def\t{\tau}       

\def\dg{\dagger}  

   \newcommand{\MeV}{ {\rm MeV} }

\newcommand{\lsp}{ \left ( } \newcommand{\rsp}{ \right ) } \newcommand{\Lg}{\mathcal{L}}  \newcommand{\To}{\Rightarrow}   \newcommand{\Getsto}{\Leftrightarrow}

\newcommand{\vev}[1]{ \langle {#1} \rangle }

  \newcommand{\Det}{{\rm Det}}

\newcommand{\Column}[3]{ \begin{pmatrix} #1 \\ #2 \\ #3 \end{pmatrix} }

\newcommand{\Diag}[3]{ \begin{pmatrix} #1 & 0 & 0 \\ 0 & #2 & 0 \\ 0 & 0 & #3 \\\end{pmatrix}}

\begin{document}

\title{\Large 
Conditions of naturalness and fine-tunings \\ for type-I seesaw mechanism with four-zero texture}

\preprint{STUPP-21-251}
\author{Masaki J. S. Yang}
\email{yang@krishna.th.phy.saitama-u.ac.jp}
\affiliation{Department of Physics, Saitama University, 
Shimo-okubo, Sakura-ku, Saitama, 338-8570, Japan}


\begin{abstract} 

In this paper, we search for conditions that the mass matrix of light neutrinos $m_{\nu}$ is not a result of large cancellations for the type-I seesaw mechanism with four-zero texture.
For the Yukawa matrix of neutrinos $Y_{\nu}$ and heavy Majorana mass matrix $M_{R}$, these conditions are written as $(Y_{\nu})_{i2} \propto (m_{\nu})_{i2} \, \Rightarrow \, (Y_{\nu})_{i2} \propto (M_{R})_{i2}$. 
We call them {\it alignment} conditions because they align the certain rows or columns of the three neutrino mass matrices.
If these conditions do not hold, the large mixing in $m_{\nu}$ is a result of fine-tuning due to the cancellation of several terms. 
Then they are required from a viewpoint of naturalness. 

They give an explanation of the seesaw-invariance of four-zero texture, and
place rough restrictions on flavor structures of neutrinos. 
Under these conditions, $Y_{\nu}$ must have a cascade hierarchy. 
For $M_{R}$, the 12 submatrix has a similar hierarchy as $Y_{\nu}$ and $m_{\nu}$. 
However, the 23 submatrix has a waterfall hierarchy without some fine-tuning.
Therefore, it is likely that $Y_{\nu}$ and $M_{R}$ have qualitatively different flavor structures.
Furthermore, since the conditions restrict CP phases of the matrix elements, 
they imply existence of a universal generalized CP symmetry in the neutrino sector.

\end{abstract} 

\maketitle

\section{Introduction}

Research on the peculiar flavor structure of the Standard Model may provide some hints to the flavor puzzle and the theoretical origin of the Higgs boson.
On the other hand, treatments of the flavor strongly depend on what we consider the Higgs boson is.
Therefore, model-independent texture studies of mass matrices $M_{f}$ are a dominant approach.

Among them, the four-zero texture 
$(M_{f})_{11} = (M_{f})_{13,31} = 0$ \cite{Fritzsch:1995nx} 
is still one of viable texture \cite{Lehmann:1995br, Kang:1997uv, Mondragon:1998gy, Chkareuli:1998sa, Nishiura:1999yt, Matsuda:1999yx, Fritzsch:1999ee, Fritzsch:2002ga, Xing:2003zd, Xing:2003yj, Bando:2004hi, Matsuda:2006xa, Branco:2006wv, Ahuja:2007vh, Barranco:2010we, Verma:2010jy, Gupta:2011zzg, Xing:2015sva, Yang:2016crz, Yang:2020qsa,  Yang:2020goc, Yang:2021smh, Yang:2021xob, Bagai:2021nsl, Fritzsch:2021ipb}. 
This system has a nice property called seesaw invariance \cite{Nishiura:1999yt, Fritzsch:1999ee} in the type-I seesaw mechanism \cite{Minkowski:1977sc,GellMann:1980v,Yanagida:1979as}. 
By imposing the four-zero texture on the right-handed neutrino mass $M_R$ and the Yukawa matrix $Y_{\n}$, the mass of light neutrinos $m_{\n}$ also becomes the four-zero texture.
However, the type-I seesaw relation of four-zero texture has many terms and its physical meaning is unclear.

Several studies have attempted to grasp the naturalness \cite{tHooft:1979bh} of the seesaw parameters \cite{Dermisek:2004tx,Sayre:2007ps,Meloni:2012sx}, 
in a meaning that there is no significant cancellation between terms. 
In this paper, we explore such conditions that no large cancellation occurs in this type-I seesaw relation with four-zero texture, their properties, and consequences.

This paper is organized as follows. 
In the next section, we explore conditions under which the seesaw relation with the four-zero texture does not have large cancellations.
Sec.~3 gives a survey on properties of such conditions and their restrictions on the flavor structure of neutrinos. 
In Sec.~4, we discuss a realization of conditions and consequences of imposing them on $m_\n$ with the trimaximal mixing condition.
The final section is devoted to a summary.

\section{Type-I seesaw mechanism and four-zero texture}

In this section, we discuss cancellations between terms of the neutrino mass matrices in the type-I seesaw mechanism with four-zero texture.
Suppose that the Dirac neutrino mass matrix $M_{D}$ and the Majorana mass matrix of right-handed neutrinos $M_{R}$ have the following four-zero texture;
\begin{align}
M_{D} = {v \over \sqrt 2} Y_{\n} = 
\begin{pmatrix}
 0 & C_{\nu } & 0 \\
 C_{\nu }^{*} & D_{\nu } & B_{\nu } \\
 0 & B_{\nu }^{*} & A_{\nu } \\
\end{pmatrix} , 
~~~ 
M_{R} = 
\begin{pmatrix}
 0 & C_R & 0 \\
 C_R & D_R & B_R \\
 0 & B_R & A_R \\
\end{pmatrix} . 
\label{Eq1}
\end{align}
The Hermiticity of $M_{D}$ can be justified by the parity symmetry of the left-right symmetric model \cite{Pati:1974yy,Senjanovic:1975rk,Mohapatra:1974hk}. 
Since the phases of $B_{\n}$ and $C_{\n}$ can be removed by redefinition of fields, 
$M_{D}$ and $Y_{\n}$ are set to be real matrices without loss of generality. 
Thus, let  $A_{\n}\sim D_{\n}$ be real parameters and $A_{R} \sim D_{R}$ be complex ones.

By the type-I seesaw mechanism, the mass matrix of light neutrinos $m_{\n}$ also have the four-zero texture \cite{Nishiura:1999yt,Fritzsch:1999ee}; 
\begin{align}
m_{\n} &= M_{D} M_{R}^{-1} M_{D}^{T} 
\\ & = 
\begin{pmatrix}
 0 & 0 & 0 \\
 0 & \frac{F_{\n}^2}{A_R} & \frac{A_{\nu } F_{\n}}{A_R} \\
 0 & \frac{A_{\nu } F_{\n}}{A_R} & \frac{A_{\nu }^2}{A_R} \\
\end{pmatrix}
+
\begin{pmatrix}
 0 & \frac{C_{\nu }^2}{C_R} & 0 \\
 \frac{C_{\nu }^2}{C_R} & \frac{2 C_{\nu } D_{\nu }}{C_R}-\frac{C_{\nu }^2 D_R}{C_R^2} & \frac{B_{\nu } C_{\nu }}{C_R} \\
 0 & \frac{B_{\nu } C_{\nu }}{C_R} & 0 \\
\end{pmatrix}
 = 
\begin{pmatrix}
 0 & c & 0 \\
 c & d & b \\
 0 & b & a \\
\end{pmatrix} ,
\label{mn} 
\end{align}
where $F_{\n} = (B_{\nu } C_R-B_R C_{\nu }) / C_{R}. $
In general, $a \sim d$ are also complex parameters.
Note that the first matrix has determinant zero and its rank is equal to one.
If $|A_{\n}|\gtrsim |F_{\n}|$ holds, the contributions of the first matrix in Eq.~(\ref{mn}) can be neglected except for the 33 element, because the parameters $a, b,$ and $d$ are expected to be similar magnitude from the bi-maximal mixing of $\th_{23}$. 
By assuming $Y_{\n}$ and $M_{D}$ have the following hierarchy, 
\begin{align}
|A_{\n}| \gg |B_{\n}|, \, |D_{\n}| , |C_{\n}| \, , 
\label{hier}
\end{align}
this requirement is rewritten as
\begin{align}
|A_{\n} | \gtrsim |F_{\n}| = \left | B_{\n} - C_{\n} {B_R \over C_{R}} \right |  ~~ \Getsto ~~ 
\left | {A_{\n} \over C_{\n}} \right |  \gtrsim \left| {B_R \over C_{R}} \right | \, . 
\label{cond0}
\end{align}
The condition~(\ref{cond0}) compares degrees of hierarchies between $M_{R}$ and $M_{D}$.
Under Eq.~(\ref{cond0}), the form of $m_{\n}$ becomes rather concise;
\begin{align}
m_{\n} & \simeq 
\begin{pmatrix}
 0 & \frac{C_{\nu }^2}{C_R} & 0 \\
 \frac{C_{\nu }^2}{C_R} &
  {C_{\n} D_{\n} \over C_R} + {C_{\n} \over C_R} (D_{\n} - C_{\n} {D_R \over C_R}) & \frac{B_{\nu } C_{\nu }}{C_R} \\
 0 & \frac{B_{\nu } C_{\nu }}{C_R} &  \frac{A_{\nu }^2}{A_R}  \\
\end{pmatrix} 
= 
\begin{pmatrix}
 0 & c & 0 \\
 c & d & b \\
 0 & b & a \\
\end{pmatrix} .
\label{mn2} 
\end{align}
In this case, we obtain
\begin{align}
 {b \over c} \simeq {B_{\n} \over C_{\n}} \, , 
 \label{casc}
\end{align}
a relation of ratios between elements of $m_{\n}$ and $Y_{\n}$.

On the other hand, when Eqs.~(\ref{cond0}) and (\ref{casc}) do not hold and the hierarchy of $M_{R}$ is so strong $| {A_{\n} / C_{\n}}| \ll | {B_R / C_{R}} |$,
the terms containing $F_{\n}$ in Eq.~(\ref{mn}) are quite larger than the other matrix elements $m_{\n}\sim C_{\n}^{2} / C_{R}\sim A_{\n}^{2} / A_{R}$. 
In such a situation, the large mixings of neutrinos are a result of fine-tuning between terms with large magnitude.
Therefore, the condition~(\ref{casc}) is also required from the naturalness in the meaning of no fine-tunings.

When the conditions (\ref{cond0}) and (\ref{casc}) hold, the smallness of $F_{\n}$ itself is also not a consequence of a large cancellation;
\begin{align}
\left | {A_{\n} F_{\n} \over A_{R}} \right | \ni \left | - { A_{\n}\over A_{R}} { C_{\n} B_R \over C_{R}} \right | \lesssim \left | {A_{\n}^{2} \over A_{R}} \right | \, . 
\label{cancel}
\end{align}
Thus, the absolute values of the terms in $A_{\n} F_{\n} / A_{R}$ are also smaller than those in $m_{\n}$, respectively. 

A similar relation holds for the 22 element of $m_{\n}$. 
From $|b| \sim |d|$ in Eq.~(\ref{mn2}), a condition for a large cancellation becomes
\begin{align}
{2 \, C_{\n} D_{\n} \over C_R} \simeq
{C_{\n}^{2} D_R \over C_R^{2}} , ~~ 
|   {C_{\n} D_{\n} \over C_R} | \gg  |   {C_{\n} B_{\n} \over C_R} | . 
\label{8}
\end{align}
This can be rewritten as the following,  
\begin{align}
\left | D_{\n} - C_{\n} {D_R \over C_R}  \right | \simeq |D_{\n}| 
 \gg | B_{\n} | \, . 
\end{align}
On the contrary, a condition for such cancellation not to occur is
\begin{align}
\left | D_{\n} - C_{\n} {D_R \over C_R}  \right | \lesssim | B_{\n} | \, . 
\label{cond2}
\end{align}
In this case $d\simeq C_{\n} D_{\n} / C_{R}$ holds and we obtain
\begin{align}
{d \over c} \simeq {D_{\n} \over C_{\n}}  \, .
\label{casc2}
\end{align}

The same condition as Eq.~(\ref{cancel}) can hold for $D_{\n}$. 
The 22 element of $m_{\n}$ is
\begin{align}
(m_{\n})_{22} = d \simeq c \,  {D_{\n} \over C_{\n}} = 
 \frac{C_{\nu } D_{\nu }}{C_R} \, ,
\end{align}
and each term in Eqs.~(\ref{mn}) and (\ref{mn2}) has an absolute value of only about $(m_{\n})_{22}$.
Therefore, Eqs.~(\ref{casc}) and (\ref{casc2}) are sufficient conditions for the naturalness. 

Note that these relations include phases of $b,c,$ and $d$.
Since $B_{\n}, C_{\n},$ and $D_{\n}$ are taken as real parameters, 
relative phases of $b, c,$ and $ d$ must be almost zero (or $\pi$) in this basis.
The phases of $A_{\n}$ and $A_{R}$ are not restricted by these conditions. If these two phases are aligned as well as the other parameters $B_{\n}, C_{\n},$ and $D_{\n}$, 
such a situation is realized by the diagonal reflection symmetries (DRS) 
 \cite{Yang:2020goc,Yang:2021smh,Yang:2021xob}: 
\begin{align}
 R \,  m_{\n}^{*} \, R  = m_{\n} , ~~~
R \,  M_{D, R}^{*} \, R = M_{D, R} ,  
\label{DRS}
\end{align}
where
\begin{align}
  R  = \Diag{-1}{1}{1}, ~~~
(m_{\n} ,  M_{D,R}) = 
\begin{pmatrix}
F_{\n,D,R} & i \, C_{\n,D,R} & i \, E_{\n,D,R} \\
 i \, C_{\n,D,R} & D_{\n,D,R} & B_{\n,D,R} \\
 i \, E_{\n,D,R} & B_{\n,D,R} & A_{\n,D,R}
\end{pmatrix} .
\end{align}
By redefining  phases of $l_{L}$ and $\n_{R}$, 
 all the mass matrices $m_{\n}$ and $M_{D,R}$ in  the neutrino sector have only real parameters, and indeed the  conditions~(\ref{casc}) and (\ref{casc2}) are real.
More generally, this situation can be achieved 
by an imposition of the same generalized CP symmetry (GCP) \cite{Ecker:1983hz, Gronau:1985sp, Feruglio:2012cw, Holthausen:2012dk} on the neutrino sector.
For this purpose, it is sufficient that the GCPs of $M_{D}$ and $M_{R}$ are the same in the seesaw mechanism,
\begin{equation}
X M_{D}^{*} X^{\dg} = M_{D} \, ,  ~~
X^{*} M_{R}^{*} X^{\dg} = M_{R} \, ,  
~~ \To ~~ 
X m_{\n}^{*} X^{T} = m_{\n} \, ,  
\end{equation}
with a Hermitian unitary matrix $X = X^{\dg}$.

\subsection{Reasons for conditions} 

Here, we discuss why such conditions are necessary.
When the mass matrices~(\ref{Eq1}) are expressed as $M_{D} = (\bm y_{1}, \, \bm y_{2} , \, \bm y_{3})$ and $M_{R} = (\bm M_{1}, \, \bm M_{2} , \, \bm M_{3})$ with three dimensional vectors $\bm y_{i}$ and $\bm M_{i}$, the inverse matrix of $M_{R}$ can be written as
\begin{align}
M_{R}^{-1} = 
{1 \over {\rm Det} \, M_{R}} 
\begin{pmatrix}
 A_R D_R - B_R^2 & -A_R C_R & B_R C_R \\
 -A_R C_R & 0 & 0 \\
 B_R C_R & 0 & -C_R^2 \\
\end{pmatrix}
= 
{1 \over {\rm Det} \, M_{R}} 
\Column
{(\bm M_{2} \times \bm M_{3})^{T}}
{(\bm M_{3} \times \bm M_{1})^{T}}
{(\bm M_{1} \times \bm M_{2})^{T}}
\label{MRinv} . 
\end{align} 
The first matrix in $m_{\n}$~(\ref{mn}) is generated from a part of $M_{R}^{-1}$;
\begin{align}
m_{\n} \ni {1 \over {\rm Det} \, M_{R}} 
\begin{pmatrix}
 0 & C_{\nu } & 0 \\
 C_{\nu } & D_{\nu } & B_{\nu } \\
 0 & B_{\nu } & A_{\nu } \\
\end{pmatrix}
\begin{pmatrix}
 -B_R^2 & 0 & B_R C_R \\
 0 & 0 & 0 \\
 B_R C_R & 0 & -C_R^2 \\
\end{pmatrix}
\begin{pmatrix}
 0 & C_{\nu } & 0 \\
 C_{\nu } & D_{\nu } & B_{\nu } \\
 0 & B_{\nu } & A_{\nu } \\
\end{pmatrix}
 = 
\begin{pmatrix}
 0 & 0 & 0 \\
 0 & \frac{F_{\n}^2}{A_R} & \frac{A_{\nu } F_{\n}}{A_R} \\
 0 & \frac{A_{\nu } F_{\n}}{A_R} & \frac{A_{\nu }^2}{A_R} \\
\end{pmatrix} \, . 
\end{align}
By comparing the 32 and 33 elements of this matrix, 
the restriction~(\ref{cond0}) on $F_{\n}$ is rewritten as a condition on the third row of $M_{R}^{-1}$ as follows 
\begin{align}
|F_{\n}| \lesssim |A_{\n}| ~~ \To ~~ 
| \Det (\bm M_{1}, \, \bm M_{2} , \, \bm y_{2}) |
\lesssim
| \Det (\bm M_{1}, \, \bm M_{2} , \, \bm y_{3}) |  \, . 
\label{16}
\end{align}
The rest of $m_\n$ comes from
\begin{align}
m_{\n} \ni {1 \over {\rm Det} \, M_{R}} 
\begin{pmatrix}
 0 & C_{\nu } & 0 \\
 C_{\nu } & D_{\nu } & B_{\nu } \\
 0 & B_{\nu } & A_{\nu } \\
\end{pmatrix}
\begin{pmatrix}
 A_R D_R  & -A_R C_R & 0 \\
 -A_R C_R & 0 & 0 \\
0 & 0 & 0 \\
\end{pmatrix}
\begin{pmatrix}
 0 & C_{\nu } & 0 \\
 C_{\nu } & D_{\nu } & B_{\nu } \\
 0 & B_{\nu } & A_{\nu } \\
\end{pmatrix}
 = 
\begin{pmatrix}
 0 & \frac{C_{\nu }^2}{C_R} & 0 \\
 \frac{C_{\nu }^2}{C_R} & \frac{2 C_{\nu } D_{\nu }}{C_R}-\frac{C_{\nu }^2 D_R}{C_R^2} & \frac{B_{\nu } C_{\nu }}{C_R} \\
 0 & \frac{B_{\nu } C_{\nu }}{C_R} & 0 \\
\end{pmatrix} \, . 
\end{align}
In a similar way, under the assumption that Eq.~(\ref{16}) holds, 
the condition (\ref{cond2}) for $D_{R}$  can be rewritten as; 
\begin{align}
\left | D_{\n} - C_{\n} {D_R \over C_R}  \right | \lesssim | B_{\n} | \
~~ \To ~~
| \Det (\bm M_{2}, \, \bm M_{3} , \, \bm y_{2}) |
& \lesssim 
| \Det (\bm M_{2}, \, \bm M_{3} , \, \bm y_{3}) |  \, . 
\end{align}
If the vectors $\bm M_{1,2,3}$ are linearly independent, 
the conditions that the two determinants are small can be rewritten as 
\begin{align}
| \Det (\bm M_{2}, \, \bm M_{3} , \, \bm y_{2}) | , \, 
| \Det (\bm M_{1}, \, \bm M_{2} , \, \bm y_{2}) |
\simeq 
0 ~~ \Getsto ~~ \bm y_{2} \simeq \a \, \bm M_{2} \simeq \b \,  \bm m_{2} \,  .
\end{align}
Here $\a , \b$ are some complex coefficients and
$m_{\n} \equiv (\bm m_{1}, \, \bm m_{2} , \, \bm m_{3})$. 
In an equivalent notation, we arrive at the following result, 
\begin{align}
(M_{D})_{i2} \simeq \a (M_{R})_{i2} \simeq \b (m_{\n})_{i2} \, . 
\label{alignment}
\end{align}
These conditions align the certain rows (or columns) of the three mass matrices of neutrinos.  
Therefore, we call Eqs.~(\ref{casc}) and (\ref{casc2}), or Eq.~(\ref{alignment}) as {\it alignment conditions}.
Similar conditions are expected to hold for a wide range of textures other than the four-zero texture. 

In the four-zero texture~(\ref{Eq1}), the first row and column satisfy the exact alignment conditions $(M_{D})_{i1} = \a (M_{R})_{i1} = \b (m_{\n })_{i1}$, which is the cause of the seesaw invariance \cite{Nishiura:1999yt, Fritzsch:1999ee}.
Since all the other seesaw-invariant four-zero textures \cite{Nishiura:1999yt} have two texture zeros in the same row or column, 
we can also consider the invariance as a result of the alignment conditions. 

The discussion here is similar to the sequential dominance \cite{King:1998jw,King:1999cm,King:1999mb,King:2002nf,Antusch:2004gf}. 
However, it is different in that $M_{R}$ is not diagonalized. 
The result obtained on the trimaximal condition later is also different from the constrained sequential dominance \cite{King:2005bj,Antusch:2011ic,Antusch:2013wn,Bjorkeroth:2014vha}.

\section{Properties of alignment conditions}

In this section, some properties of the alignment conditions are explored. 
First of all, we will consider parameter region where these conditions (\ref{casc}) and (\ref{casc2}) hold.
The Majorana mass matrix of $\n_{R}$ is represetned by $M_{D}$ and $m_{\n}$ as
\begin{align}
M_{R} &= M_{D}^{T} \, m_{\n}^{-1} \, M_{D} \\ & = 
\begin{pmatrix}
 0 & 0 & 0 \\
 0 & \frac{F_{R}^{2}}{a} & \frac{A_{\nu } F_{R}}{a} \\
 0 & \frac{A_{\nu } F_{R}}{a} & \frac{A_{\nu}^2}{a} \\
\end{pmatrix} + 
\begin{pmatrix}
 0 & \frac{C_{\nu }^2}{c} & 0 \\
 \frac{C_{\nu }^2}{c} & {C_{\n} \, D_{\n} \over c} + {C_{\n}  \over c} (D_{\n} - {d \over c} C_{\n} )& \frac{B_{\nu } \,C_{\nu }}{c} \\
 0 & \frac{B_{\nu } \, C_{\nu }}{c} & 0 \\
\end{pmatrix} \, 
= 
\begin{pmatrix}
 0 & C_R & 0 \\
 C_R & D_R & B_R \\
 0 & B_R & A_R \\
\end{pmatrix} , 
\label{MR}
\end{align}
where $F_{R} = (c \, B_{\nu} - b \, C_{\nu})/ c$. 
Imposing the inequality (\ref{cond0}) on the 12 and 23 elements of $M_{R}$, we obtain
\begin{align}
\left| \frac{B_R }{C_{R}} \right | &= 
\left|
\frac{A_{\nu } \, (c \, B_{\nu} - b \, C_{\nu})+ a B_{\nu } \, C_{\nu } }{a \, C_{\n}^{2}} 
\right | 
\lesssim \left | {A_{\n} \over C_{\n}} \right |  \, . 
\label{24}
\end{align}
Since the term $a B_{\n} C_{\n}$ can be neglected from the hierarchy (\ref{hier}), 
Eq.~(\ref{24}) restricts a deviation $\e$ of $B_{\n}$ from Eq.~(\ref{casc}) as
\begin{align}
B_{\n} \equiv  {b \over c}  C_{\nu} + \e, ~~~
| \e | = \left | B_{\nu} - {b \over c}  C_{\nu}  \right |  
\lesssim \left |  \frac{a \, C_{\nu }}{c } \right |  .
\label{dev1}
\end{align}
Thus, this $\e$ is restricted to a region with a width of about $C_{\n}$.
For example, by assuming $Y_{\n}\simeq Y_{e}$, the range of this limit is about $C_{e}\simeq \sqrt{m_{\m} m_{e}}\sim 7\, \MeV$ with $B_{e}\simeq m_{\m}\sim 100\, \MeV$. The range is sufficiently large to allow renormalizations and threshold corrections.
From $|a| \sim |b|$, the magnitude of $B_{\n}$ for this range of $|\e|$ becomes $0 \lesssim | B_{\n} | \lesssim 2 b \, C_{\n} / c$.

Similarly,  we define a complex parameter $\d$ for the 22 element; 
\begin{align}
D_{\n} \equiv  {d \over c}  C_{\nu} + \d .
\label{dev2}
\end{align}
Eq.~(\ref{cond2}) and (\ref{MR}) leads to
\begin{align}
\left| D_{\n} -  C_{\n} {D_R \over C_R} \right | 
= \left| - ( D_{\n} - {d \over c} C_{\n})  \right | 
= | \d| \lesssim |B_{\n}| \simeq \left | {b \, C_{\n} \over c} \right | \, . 
\label{19}
\end{align}
From $|a|\sim |b|\sim |d|$, the allowed ranges of $\e$ and $\d$ are comparable.

\subsection{Flavor structure and alignment conditions}

The alignment conditions place rough constraints on the hierarchy of $Y_{\n}$ and $M_{R}$. 
If the conditions (\ref{casc}) and (\ref{casc2}) hold, 
the hierarchy of $Y_{\n}$ becomes a cascade-type. 
Here, waterfall (geometric) and cascade texture are matrices of the following form \cite{Dorsner:2001sg, Haba:2008dp}.
\begin{align}
M_{\rm waterfall} \sim 
\begin{pmatrix}
\d^{2} & \l \, \d & \d  \\
\l \, \d  & \l^{2} & \l \\
\d & \l & 1 \\
\end{pmatrix} ,
~~~
M_{\rm cascade} \sim 
\begin{pmatrix}
\d & \d & \d \\
\d & \l & \l \\
\d & \l & 1 \\
\end{pmatrix} .
\label{watercas}
\end{align}
Although the paper \cite{Haba:2008dp} assumed $1 \gg \l \gg \d$, this paper includes a situation with $\l \gtrsim \d$. 
If the four-zero texture arises from sequential breaking of flavor symmetry such as $SU(3)_{F}$ (or its subgroup), the cascade type is desirable for flavor structures of $Y_{\n}$ and $M_{D}$.
The observation of the CKM matrix and some unification also suggest cascade type for Hermitian $Y_{d}$ and $Y_{e}$ \cite{Yang:2021smh}.

Furthermore, the bi-maximal mixing of $\th_{23}$ implies $|a|\sim |b|\sim |d|$ in $m_{\n}$ (\ref{mn2}).
The bi-maximality places constraints on magnitudes of matrix elements. First, the condition $|a| \sim |b|$ in Eq.~(\ref{mn2}) leads to 
\begin{align}
 {B_{\n} C_{\n} \over C_{R}}  \sim {A_{\n}^{2} \over A_{R}} 
~~ \Getsto ~~ 
{A_{R} \over C_{R}} \sim  {A_{\n} \over B_{\n}} {A_{\n} \over C_{\n}} .
\label{rel1}
\end{align}
Thus, the magnitude of $M_{R}$ for the third generation is enhanced by $A_{\n} / B_{\n}$ and much larger than that of $Y_{\n}$.
If there is also no fine-tunings between the terms in the 22 element,  $d \sim b$ also leads to 
\begin{align}
 {D_{R} \over  C_{R} } \sim {D_{\n} \over C_{\n}} \sim {B_{\n} \over C_{\n}} \simeq {b \over c} .
 \label{rel2}
\end{align}
This suggests a mild hierarchical structure for light generations of $M_{R}$, similar to that of $Y_{\n}$ and $m_{\n}$.

These facts can also be seen from the reconstruction of $M_{R}$.
Substituting Eqs. (\ref{dev1}) and (\ref{dev2}) into Eq. (\ref{MR}), $M_{R}$ for given $Y_{\n}$ and an upper limit of its absolute value are 
\begin{align}
M_{R} = 
\begin{pmatrix}
 0 & \frac{C_{\nu }^2}{c} & 0 \\
 \frac{C_{\n}^2}{c} 
 & {\e^2 \over a} + {d \, C_{\n}^2 \over c^2} + {2 \, \d \, C_{\n}  \over c} & \frac{\e  A_{\nu }}{a}+\frac{b \, C_{\n}^2}{c^2}+\frac{\e \, C_{\n}}{c} \\
 0 & \frac{\e  A_{\n}}{a}+\frac{b \, C_{\n}^2}{c^2}+\frac{\e \, C_{\n}}{c} & \frac{A_{\nu }^2}{a} \\
\end{pmatrix}
\lesssim 
{C_{\n} \over c}
\begin{pmatrix}
 0 & C_{\nu}  & 0 \\
 C_{\n} & 4 \, D_{\n} & A_{\nu }  \\
 0 &  A_{\nu }  & \frac{c \, A_{\nu }^2}{a \, C_{\n}} \\
\end{pmatrix} . 
\label{MR2}
\end{align}
In this case, $M_{R}$ has a partial cascade hierarchy and satisfies Eqs.~(\ref{rel1}) and ({\ref{rel2}}).
On the other hand, in a situation like $|B_{\n}| , |D_{\n}| \gg |b \, C_{\n} / c|$, Eq.~(\ref{rel1}) and Eq.~(\ref{rel2}) do not hold and $M_{R}$ will be a waterfall type matrix  \cite{Yang:2016crz}; 
\begin{align}
M_{R} \simeq 
\begin{pmatrix}
 0 & \frac{C_{\nu }^2}{c} & 0 \\
 \frac{C_{\nu }^2}{c} & \frac{B_{\nu }^2}{a} + { 2 D_{\n} C_{\n} \over c} & \frac{A_{\nu } B_{\nu }}{a} \\
 0 & \frac{A_{\nu } B_{\nu }}{a} & \frac{A_{\nu }^2}{a} \\
\end{pmatrix} \, . 
\end{align}

In particular, if equalities hold for the conditions~(\ref{casc}) and (\ref{casc2}) with $\e = \d = 0$, the conditions also hold for $M_{R}$~(\ref{MR2}),
\begin{align}
{D_{\n} \over C_{\n}} = {d \over c}  
~~ \Getsto ~~ 
{D_{R} \over C_{R}} = {D_{\n} \over C_{\n}} .
\label{casc4} \\
{B_{\n} \over C_{\n}} = {b \over c}  
~~ \Getsto ~~ 
{B_{R} \over C_{R}} = {B_{\n} \over C_{\n}} .
\label{casc3} 
\end{align}
Equivalently, the structure of $M_{R}$ becomes 
\begin{align}
M_{R} = {C_{\n} \over c}
\begin{pmatrix}
 0 & C_{\n} & 0 \\
 C_{\n} & D_{\n}  & B_{\n} \\
 0 & B_{\n} & {c \over a} {A_{\n}^{2} \over C_{\n}} \\
\end{pmatrix} \, 
= 
\begin{pmatrix}
 0 & C_R & 0 \\
 C_R & D_R & B_R \\
 0 & B_R & A_R \\
\end{pmatrix} . 
\end{align}
In this case, a partially universal texture of the lighter generations 
 emerges in the neutrino sector $m_{\n}, M_{D}$ and $M_{R}$. 
It is rather attractive from a viewpoint of model building.

\subsection{The conditions for $M_{R}$}

Let us investigate the range where these conditions (\ref{casc4}) and (\ref{casc3}) 
 hold for $D_{R}$ and $B_{R}$. First, by taking a ratio of the 12 and 22 elements in Eq.~(\ref{MR2}), 
\begin{align}
{(M_{R})_{22} \over (M_{R})_{12}} = 
 {d \over c} + {2 \, \d \over C_{\n}} + {\e^2 c \over a \, C_{\n}^{2}} \, .
\label{M22}
\end{align}
From Eqs.~(\ref{dev1})\,-\,(\ref{19}), $|\d|, |\e| \lesssim | b \, C_{\n} / c|$ holds. 
Then the upper limit of the absolute value of Eq.~(\ref{M22}) in this range will be about $4 d /c$, which is comparable to the allowed range $0 \lesssim |D_{\n} / C_{\n}| \lesssim 2 d / c$ for $D_{\n} / C_{\n}$.

However, the condition for $B_{R}$~(\ref{casc3}) holds in a very narrow region.
In the same way in  Eq.~(\ref{M22}), 
the ratio of matrix elements of $M_{R}$~(\ref{MR2}) becomes 
\begin{align}
{(M_{R})_{23} \over (M_{R})_{12}} = 
\frac{b}{c}+\frac{\e}{C_{\n}}+ \frac{\e \, c \, A_{\nu }}{a \, C_{\n}^{2}} .
\label{M23}
\end{align}
From this, the range of $\e$ for which Eq.~(\ref{casc3}) is approximately valid is
\begin{align}
\left |\e \, { c \, A_{\nu } \over a \, C_{\n}^{2}}  \right | \lesssim {b \over c}
~~ \To ~~ 
\left | \e   \right | \lesssim  { a \, b \, C_{\n}^{2} \over  c^{2} \, A_{\nu }}  \, . 
\label{constraint}
\end{align}
It means that the magnitude of $\e$ is constrained to about $C_{\n}^{2} / A_{\n}$.
For example, a unified relation $Y_{\n}\simeq Y_{e}$ yields a very narrow range,  $C_{e}^{2} / A_{e}\simeq m_{e} m_{\m} / m_{\t}\simeq 0.03 \, \MeV$.
Such a strict restriction is not realistic for renormalizations and or threshold corrections. 
Conversely, if we consider the upper limit of $\e$ as Eq.~(\ref{dev1}), the third term in  Eq.~(\ref{M23}) for $\e \sim C_{\n}$ is as large as $A_{\n} / C_{\n}$. 
Therefore, the condition~(\ref{casc3}) for $B_{R}$ seems to be less rigid than that of $Y_{\n}$.

In conclusion, the flavor structure of $M_{R}$ with the four-zero texture and alignment conditions can be similar to $Y_{\n}$ for the 12 submatrix. 
By contrast, it is difficult to make the same structures for the 23 submatrix of $M_{R}$ and $Y_{\n}$ without fine-tunings.
As you can see in Eq.~(\ref{MR2}), $Y_{\n}$ and $M_{R}$ are likely to have qualitatively different flavor structures.
However, even if $B_{\n}$ at low energy does not satisfy the strict limit~(\ref{constraint}), it is possible that the universality~(\ref{casc3}) ${B_{R} / C_{R}} = {B_{\n} / C_{\n}} $ is valid in a high energy region such as the GUT scale.
 
\section{Realization and application of the conditions}

In this section, we discuss a field theoretical realization of such alignment conditions and apply them to realistic textures.
Let us consider fields that transform under gauge symmetry and $SU(3)_{F} \times U(1)$ flavor symmetry, as shown in Table I.
\begin{table}[h]
  \begin{center}
    \begin{tabular}{|c|cccc|} \hline
           & $SU(2)_{L}$ & $U(1)_{Y}$ & $SU(3)_{F} $ & $U(1)$ \\ \hline \hline
      $l_{Li}$ & \bf 2 & $-1/2$ & $\bm 3$ & $+1$ \\
      $\n_{Ri}$ & \bf 1 & $0$ & $\bm 3^{*}$ & $-1$ \\ 
      $H$ & \bf 2 & $1/2$ & $\bm 1$ & $0$  \\ 
      $\D$ & \bf 1 & $0$ & $\bm 1$ & $0$ \\ \hline  
      $\h$ & \bf 1 & $1$ & $\bm 3$ & $+1$ \\ 
      $\th$ & \bf 1 & $1$ & $\bm 6$ & $+2$ \\ \hline
    \end{tabular}
    \caption{Charge assignments of fields under gauge and flavor symmetries. }
  \end{center}
\end{table}
In the context of $SU(3)_{F}$ \cite{Berezhiani:1985in,King:2001uz,Chkareuli:2001dq,King:2003rf}, a three-dimensional representation of flavon $\h$ is used to generate a four-zero texture. However, this is an antisymmetric representation and cannot be used for $M_{R}$. 
Then, a six-dimensional representation of flavon $\th$ is adopted.
The $U(1)$ symmetry prohibits the $\bf 3 \, ({\bf 6})$ representation $\h \, (\th)$ from being used an odd (even) number of times\footnote{From an irreducible decomposition of $SU(3)$, $6 \times 6 = \bar 6_{s} + 15_{a} + 15_{s}' \, , ~ 6 \times \bar 6 = 1 + 8 + 27$  \cite{Slansky:1981yr},  it is possible to write an invariant interaction using (the complex conjugate of) the six representation twice.}.

Under these symmetries, 
the general invariant Yukawa interactions up to dimension six are
\begin{align}
\Lg &= - \lsp  y_{\n}^{\h} {\h_{i} \h_{j} \over \L^{2}} + y_{\n}^{\th} {\th_{ij} \over \L} \rsp \bar l_{L i} \n_{R j} \tilde H 
- \lsp y_{R}^{\h} {\h_{i} \h_{j} \over \L^{2}} + y_{R}^{\th} {\th_{ij} \over \L} \rsp \bar \n_{Ri}^{c} \n_{Rj} \D + {\rm h.c.} \, ,
\end{align}
where $\L$ is a cut-off scale. 
The general vacuum expectation values (vevs) for the three and six representations are
\begin{align}
\vev{\h} = \Column{C}{B}{A} \, ,  ~~~
\vev{\th} = 
\begin{pmatrix}
f' & e' & d' \\
e' & c' & b' \\
d' & b' & a' \\
\end{pmatrix} \, , 
\end{align}
with complex values $A \sim C$ and $a \sim f$. 

By redefining the neutrino fields (or weak basis transformations (WBT) \cite{Branco:1999nb,Branco:2007nn}), it does not lose generality 
to rewrite $\vev{\h} = (0, \, 0 , \, A)^{T}$. 
Furthermore, the 11 element of $\vev{\th}$ can be set to zero by a WBT for 12 mixing. 
It leads to 
\begin{align}
\vev{\h} = \Column{0}{0}{A} \, ,  ~~~
\vev{\th} = 
\begin{pmatrix}
0 & e  & d \\
e & c & b \\
d & b & a \\
\end{pmatrix} \, . 
\end{align}
For hierarchical flavor structures, we assume that $y_{\n, R}^{\h} A^{2} / \L^{2} \gg y_{\n, R}^{\th} \vev{\th}_{ij} / \L$ in this basis.

As $\D$ and $H$ acquire vevs, the neutrino mass matrices are respectively
\begin{align}
\tilde M_{D} = {y_{\n}^{\th} \over \L} \vev{H}
\begin{pmatrix}
0 & e  & d \\
e & c & b \\
d & b & a + {y_{\n}^{\h} \over y_{\n}^{\th}} {A^{2} \over \L} \\
\end{pmatrix} \, , ~~~
\tilde M_{R} = {y_{R}^{\th} \over \L} \vev{\D}
\begin{pmatrix}
0 & e  & d \\
e & c & b \\
d & b & a + {y_{R}^{\h} \over y_{R}^{\th}} {A^{2} \over \L} \\
\end{pmatrix} \, , 
\end{align}
In this basis, the first and second generations satisfy the exact alignment conditions;
\begin{align}
(\tilde M_{D})_{i1} = \a (\tilde M_{R})_{i1} = \b (m_{\n })_{i1}\, , ~~~ 
(\tilde M_{D})_{i2} = \a' (\tilde M_{R})_{i2} = \b' (m_{\n })_{i2}\, . 
\end{align}
Furthermore, when the 13 and 31 elements are set to zero by WBTs of 23 mixing with $\tan \phi = d / e$, a four-zero texture with broken alignment is realized. 
The 22 elements of mass matrices in such a basis are 
\begin{align}
(\tilde M_{D})_{22} &= c \, \cos^{2} \phi + 2 b \, \sin \phi \cos \phi + \lsp a + {y_{\n}^{\h}\over y_{\n}^{\th}} {A^{2}\over \L} \rsp \sin^{2}\phi \, , \\
(\tilde M_{R})_{22} &= c \, \cos^{2} \phi + 2 b \, \sin \phi \cos \phi + \lsp a + {y_{R}^{\h}\over y_{R}^{\th}} {A^{2}\over \L} \rsp \sin^{2}\phi \, . 
\end{align}
Breaking effects are comes from the difference in the ratio of ${y_{\n}^{\h} / y_{\n}^{\th}}$ and ${y_{R}^{\h} / y_{R}^{\th}}$.
However, if the mixing angle $\phi$ is small (i.e., $d \ll e$), the effects  ${y_{\n , R}^{\h}\over y_{\n,R}^{\th}} {A^{2}\over \L}\sin^{2}\phi $ can also be small.
For example, 
${y_{\n,R}^{\h}\over y_{\n,R}^{\th}} {A^{2}\over \L} : c \simeq m_{t} : m_{c} \simeq 300 : 1$ holds like the up-type quarks, conditions to prevent a misalignment 
$| {y_{\n,R}^{\h}\over y_{\n,R}^{\th}} {A^{2}\over \L}\sin^{2}\phi | \lesssim c$ leads to 
an allowed region of $\phi$ as $|\sin \phi| \lesssim \sqrt{1 / 300} \simeq 0.06$. 
By constrast, if ${y_{\n,R}^{\h}\over y_{\n,R}^{\th}} {A^{2}\over \L} : c \simeq 16 : 1$ holds as in the charged leptons, 
the allowed region of $\phi$ becomes $|\sin \phi| \lesssim \sqrt{1 / 16} \simeq 0.25$ and it has a wide parameter range.

Although the effect on the $23$ element $b$ is about $ {y_{R}^{\h} \over y_{R}^{\th}} {A^{2} \over \L} \sin \phi \cos \phi$, the alignment condition does not have to be so rigid by Eq.~(\ref{M23}).
In this way, the alignment condition~(\ref{casc4}) is not difficult to achieve
 if there is a flavon that  generates the first and second generation of $M_{D}$ and $M_{R}$ simultaneously.

\subsection{Application to trimaximal mixing and magic symmetry}

Here we will analyze a neutrino mass matrix with four-zero texture that satisfies the alignment conditions and the trimaximal mixing \cite{Harrison:2002kp,Harrison:2004he,Friedberg:2006it,Lam:2006wy,Bjorken:2005rm, He:2006qd, Grimus:2008tt}. 
The four-zero texture with a trimaximal condition is given by 
 \cite{Gautam:2016qyw}
\begin{align}
m_{\n T} = 
\begin{pmatrix}
0 &  c & 0 \\
 c & c+a & c+a \\
0 & c+a & a
\end{pmatrix} , 
~~~ 
m_{\n T} \Column{1}{-1}{1} = -c \Column{1}{-1}{1} . 
\label{mnT}
\end{align}
This matrix satisfies the following magic symmetry \cite{Lam:2006wy}, which is equivalent to the eigenvector condition~(\ref{mnT});
\begin{align}
S_{2} \, m_{\n T} \, S_{2} = m_{\n T}, ~~~ 
S_{2} = 
\begin{pmatrix}
 \frac{1}{3} & \frac{2}{3} & -\frac{2}{3} \\[2pt]
 \frac{2}{3} & \frac{1}{3} & \frac{2}{3} \\[2pt]
- \frac{2}{3} & \frac{2}{3} & \frac{1}{3} \\
\end{pmatrix} ,
~~  S_{2}^{2} = 1_{3} \, . 
\label{magic}
\end{align}
However, even if we impose the trimaximal condition (\ref{mnT}) on the general $m_{\n}$~(\ref{mn}), there are so many terms and physical meanings are unclear.
Besides, if the trimaximal mixing is a consequence of fine-tunings, 
it is difficult to derive relations for the matrix elements of $Y_{\n}$ and $M_{R}$. 

Therefore, here we impose the alignment conditions~(\ref{alignment}), 
or equivalently Eqs.~(\ref{casc}) and (\ref{casc2}).
Similar results have been obtained in several studies \cite{Dermisek:2004tx,Sayre:2007ps, Haba:2008dp}. Relations $B_{\n} / D_{\n} = B_{R} / D_{R} = 1$ are relatively easy to realize in a $A_{4}$ model with a flavon with vev $\vev{\varphi} = ( 0, 1,1)$. 
A specific realization can be seen in Ref.~\cite{Yang:2021xob}.  
Although the other one $C_{\n} / D_{\n} = C_{R} / D_{R}$ is non-trivial, 
it is possible if we require some kind of unification between $l_{l}$ and $\n_{R}$ (or $Y_{\n}$ and $M_{R}$) when constructing the flavor structure. 

From Eqs.~(\ref{dev1}) and (\ref{dev2}), or 
$b = {c \over C_{\n}} ( B_{\n} - \e)$ and $d = {c \over C_{\n}} (D_{\n} - \d)$,  the mass matrix $m_{\n}$ (\ref{mn}) is represented by the deviations $\e$ and $\d$ as
\begin{align}
m_{\n} 
 & = 
\begin{pmatrix}
 0 & \frac{C_{\nu }^2}{C_R} & 0 \\
 \frac{C_{\nu }^2}{C_R} & \frac{C_{\nu }}{C_R} (D_{\n} - \d)
 & \frac{C_{\nu }}{C_R} (B_{\n} - \e) \\
 0 & \frac{ C_{\nu }}{C_R} ( B_{\n}  - \e) & \frac{A_{\nu }^2}{A_R} \\
\end{pmatrix}
 = 
\begin{pmatrix}
 0 & c & 0 \\
 c & d & b \\
 0 & b & a \\
\end{pmatrix} .
\label{mn3} 
\end{align}
By identifying Eq. (\ref{mn3}) with $m_{\n T}$~(\ref{mnT}),  two new relations arise;
\begin{align}
c + a = 
{C_{\n}^{2} \over C_{R}} + {A_{\n}^{2} \over A_{R}} 
=  {C_{\n} \over C_R} (B_{\n} - \e)  = {C_{\n} \over C_R} (D_{\n} - \d) \, . 
\end{align}
Or, equivalently, 
\begin{align}
C_{\n} \lsp 1 + {a \over c} \rsp  = D_{\n} - \d =  B_{\n} - \e \, .
\end{align}
By the alignment conditions $(Y_{\n})_{2i} \simeq \a (m_{\n})_{2i}$, the second row (or column) $(Y_{\n})_{2 i}$ also satisfies the trimaximal condition approximately. 
Without the conditions, it is difficult to extract such a relation from only the trimaximal condition (\ref{mnT}).

Under the alignment conditions, the form of $M_{D}$ is found to be
\begin{align}
M_{D} =
\begin{pmatrix}
 0 & C_{\nu } & 0 \\
 C_{\nu } &  \frac{(a+c) \, C_{\n}}{c} + \d & \frac{(a+c) \, C_{\n}}{c} + \e \\
 0 & \frac{(a+c) \, C_{\n}}{c} + \e & A_{\nu } \\
\end{pmatrix} . 
\label{MD4}
\end{align}
By further imposing DRS~(\ref{DRS}), $m_{\n T}$ (\ref{mnT}) reproduces the MNS matrix with an accuracy of $O(10^{-2})$ and predicts $a \simeq 2c$ \cite{Yang:2021xob}. 
In this case, these parameters are evaluated to $B_{\n}\sim D_{\n}\sim 3\, C_{\n}$.
A reconstructed $M_{R}$ (\ref{MR2}) from $m_{\n}$ and $M_{D}$ becomes
\begin{align}
M_{R} = 
\begin{pmatrix}
 0 & \frac{C_{\nu }^2}{c} & 0 \\
 \frac{C_{\n}^2}{c} 
 & {\e^2 \over a} + {(a + c) \, C_{\n}^2 \over c^2} + {2 \, \d \, C_{\n}  \over c} & \frac{\e  A_{\nu }}{a}+\frac{ (a + c) \, C_{\n}^2}{c^2}+\frac{\e \, C_{\n}}{c} \\
 0 & \frac{\e  A_{\n}}{a}+\frac{ (a+c) \, C_{\n}^2}{c^2}+\frac{\e \, C_{\n}}{c} & \frac{A_{\nu }^2}{a} \\
\end{pmatrix} \, . 
\label{MR4}
\end{align}
Thus, we expect a universal mild hierarchy for the lighter generations among $m_{\n}, Y_{\n}$ and $M_{R}$.

Furthermore, relations $A_{\n, R} + C_{\n,R} = B_{\n ,R}$ lead to 
the magic symmetry~(\ref{magic}) for $M_{D}$ and $M_{R}$, 
and matrix elements of $m_{\n}, M_{D}$ and $M_{R}$ are concisely related as the previous study \cite{Yang:2021xob}. 
In this case, at the cost of losing the hierarchy~(\ref{hier}), the alignment conditions~(\ref{casc}) and (\ref{casc2}) are not necessarily needed. 

Perhaps a solution $F_{\n} = A_{\n},$ or equivalently, $\e \, C_{\n} / C_{R} = a$ is similar to the sequential dominance 
\cite{King:1998jw,King:1999cm} and seems to be natural because it separates the contributions of the two mass scales $a$ and $c$ in Eq.~(\ref{mnT}). 
In this case, $B_{\n} = C_{\n}$ holds and further imposition of $B_{\n} = D_{\n}$ leads to $D_{R} = C_{R}$.
This is also similar to the constrained sequential dominance (CSD) \cite{King:2005bj}.

\section{Summary}

In this paper, we search for conditions that the mass matrix of light neutrinos $m_{\nu}$ is not a result of large cancellations between terms for the type-I seesaw mechanism 
with the four-zero texture.
For the Yukawa matrix of neutrinos $Y_{\nu}$ and heavy Majorana mass matrix $M_{R}$, these conditions have the form $(Y_{\nu})_{i2} \propto (m_{\nu})_{i2} \, \To \, (Y_{\nu})_{i2} \propto (M_{R})_{i2}$. 
We call them {\it alignment} conditions because they align the certain rows or columns of the three neutrino mass matrices.
If these conditions do not hold, the large mixings of $m_{\nu}$ is a result of fine-tunings due to the cancellation of terms with large magnitudes.
Then they are required from a viewpoint of naturalness. 
Similar conditions are expected to hold for a wide range of textures other than the four-zero texture. 
In particular, if there are two zero textures in the same row and column of $Y_{\n}$ and $M_{R}$, the mass of the light neutrinos $m_{\n}$ has the same property.  Some kind of seesaw-invariant zero textures can also be explained from the conditions. 

They place rough restrictions on flavor structures of neutrinos. 
Under these conditions, $Y_{\nu}$ must have a cascade hierarchy. 
For $M_{R}$, the 12 submatrix has a similar hierarchy as $Y_{\nu}$ and $m_{\nu}$. 
However, the 23 submatrix has a waterfall hierarchy without some fine-tuning for matrix elements of $Y_{\n}$.
Therefore, it is likely that $Y_{\nu}$ and $M_{R}$ have qualitatively different flavor structures.
Furthermore, since the conditions incorporate $CP$ phases of the matrix elements, 
they imply existence of a universal generalized $CP$ symmetry in the neutrino sector.

If the conditions are satisfied, information of $Y_{\nu}$ can be immediately extracted from a reconstructed $m_{\nu}$.
By combining the alignment conditions with the trimaximal mixing condition, 
the second row (or column) $(Y_{\n})_{2 i}$ also satisfies the trimaximal condition. 
It yields two relations
$(Y_{\nu})_{22} \simeq (Y_{\nu})_{23} \simeq 
(Y_{\nu})_{12} [(m_{\nu})_{12} + (m_{\nu})_{33}]  / (m_{\nu})_{12}$. 
Without the conditions, it is generally difficult to translate the trimaximal condition into such a relation for matrix elements in the type-I seesaw mechanism.

\section*{Acknowledgement}

This study is financially supported 
by JSPS Grants-in-Aid for Scientific Research
No.~18H01210, No. 20K14459,  
and MEXT KAKENHI Grant No.~18H05543.


\end{document}